\documentclass[pra,twocolumn,floatfix,longbibliography]{revtex4-2} 

\usepackage{amsmath}  
\usepackage{amsfonts} 
\usepackage{graphicx} 
\usepackage{xcolor}
\usepackage{tikz}

\newif\ifredlined

\redlinedtrue

\ifredlined
\newcommand\circled[1]{\marginpar{\revcolor{red}\tikz[baseline=(char.base)]{\node[shape=circle,draw,inner sep=2pt] (char) {#1};}}}
\newcommand\revcolor[1]{\color{#1}}
\else
\newcommand\circled[1]{}
\newcommand\revcolor[1]{}
\fi

\newcommand{\be}{\begin{equation}}
\newcommand{\ee}{\end{equation}}
\newcommand{\ba}{\begin{align}}
\newcommand{\ea}{\end{align}}

\newcommand\kk{\mathbf{k}}
\newcommand\kt{\mathbf{k}_t}

\begin{document}

\title{Propagation and material-interface effects in the higher-order harmonic radiation from solid-state samples.}
\author{M.~Kolesik}

\affiliation{James Wyant College of Optical Sciences, University of Arizona, Tucson, AZ 85721, U.S.A. }

\begin{abstract}
  The propagation-effects reshaping the excitation pulse are known to exhibit a strong influence on the
  high-harmonic generation (HHG) in solid-state media.
  Previous measurements showed that the mid-infrared pulse dynamics, most importantly the nonlinear
  loss and spectral broadening, can dampen or even extinguish the highest harmonic peaks.
  Despite the importance of these effects, their inclusion in the HHG-modeling has been so far restricted
  to one-dimensional propagation and/or very thin samples.
  This work  demonstrates an approach  where the driving pulse is simulated with a full spatial
  and temporal resolution in samples of realistic thickness while the material interfaces are included as well.
  We show that the HHG spectrum measured in the transmission geometry is greatly affected by the Fresnel reflections
  causing interference in the vicinity of the material surface, and we find that different parts of the harmonic
  spectra originate from different regions of the material sample.
  Our results underline the importance of realistic and comprehensive simulations in the interpretation 
  of high-harmonic generation from solids in the transmission-geometry.
\end{abstract}

\maketitle

\section{Introduction}

Higher-order harmonic generation (HHG) from solids~\cite{BrabecRev,ReisReviewHHG,recentHHGtrends}
is an extreme nonlinear effect caused by non-resonant optical excitation at high
field-intensities. While it is very much a universal behavior 
found in many systems as long as they are strongly excited away from resonance,
the underlying dynamics also reflects the properties of the
given material. This is why the  observations of the above-the-gap harmonic generation
from a solid-state medium~\cite{Ghimire11} motivated research to understand HHG as
an experimental probe. For example, a proof of principle was demonstrated early~\cite{ReconstAllOptBands}
for the all-optical reconstruction of the electronic band-structure using the measured HHG-spectra,
and similar-in-spirit investigations continue~\cite{Lanin:17,SmirnovaIvanov} to attract interest.
Beyond the energy-bands, HHG was identified as a suitable probing mechanism also
for the light-coupling transition-dipole moments~\cite{ReconstDipole,RetrievalD}, and as a way to study
the topological properties of materials~\cite{Baykusheva,PhysRevB.106.205422} and
special features of the material's electronic structure~\cite{Brabec,TunableHuber}.

The fact that HHG is so sensitive to every aspect of the microscopic dynamics in a solid
suggests that the HHG-based spectroscopy may be an approach with many different modalities.
On the other hand, precisely because the HHG measurements respond to so many properties at the same time,
it is not trivial to disentangle the observations. For example, if a thicker sample produces
a much weaker harmonic band, is the reason to be attributed solely to the energy loss of the
driving pulse, is it rather due to absorption of the harmonic radiation in the material, or perhaps
both mechanisms are important? This is the reason why numerical modeling becomes crucial.

However, numerical simulations involving HHG~\cite{Gaarde22,HHGmethods} are still rather challenging.
One of the issues is that a measured HHG spectrum is a manifestation of an interplay between the
microscopic nonlinear response and macroscopic propagation~\cite{Gaarde22} of the mid-infrared and high-harmonic
fields.  The numerical complexity required for a comprehensive simulation is probably the main reason
the modeling of solid-state HHG often concentrates on what can be dubbed a ``point-response,''
which is calculated at a single location in the material under the assumption that the local
electric field is known. The spectrum of the current density simulated at a single point is then
taken to represent the measured HHG spectrum. Obviously, this is a rather poor approximation,
especially in the transmission geometry.

When the solid-state high-harmonic spectra are generated in the transmission geometry, the 
measured spectrum is greatly affected by the dynamics of the driving pulse.
The most important mechanism, identified e.g. in~\cite{Xia:18} in experiments on GaAs,
is the high intensity of the driver pulse causing carrier generation. The associated losses
can significantly decrease the energy of the pulse by the time it reaches the exit facet of the sample,
from where most of the high-harmonic radiation is generated. As a result, the higher
harmonic orders  are  effectively ``extinguished.'' The self-phase modulation
is also significant  and it causes spectral broadening of all high-harmonic peaks.
For these reasons, the reflection geometry offers a more direct probe of the microscopic material
dynamics~\cite{ReflGeom1,Kozak23,Kozak24}. Nevertheless, it is important to understand the requirements
for an accurate interpretation of the transmission-geometry high-harmonic spectra (THHG).
Comprehensive modeling, which includes the simulation of the excitation pulse, can be
instrumental in this.

Extending the numerical simulation of HHG  beyond the ``point-model'' level presents a formidable challenge.
While it has been understood that the geometry of the experiment~\cite{Xia:18} together with
the propagation effects~\cite{Multiscale,Kilen20} play a role in shaping the measured HHG,
comprehensive simulations based on first principles, being numerically very
expensive~\cite{Multiscale,PhysRevB.107.035132,Uemoto22}, are usually reduced to one-dimensional
pulse propagation and/or are only applied to very thin material slabs.
However, under highly nonlinear conditions and/or in a realistically thick sample (hundreds of microns),
the propagation dynamics couples temporal and spatial degrees of freedom of the excitation pulse
and the one-dimensional approximation becomes insufficient.

The goal of the present work is two-fold. First, from the computational standpoint, we present
a scheme which makes it feasible to create a comprehensive model of an experiment
in the transmission geometry. The approach accurately treats the propagation effects in both the mid-infrared
excitation pulse and in the high-frequency radiation, all the while the Fresnel reflections from the
internal material boundaries are also included in the simulation. Such calculations can be done with full
spatial resolution, and are therefore capable to capture spatio-temporal coupling in ultra-short duration pulses,
which is always significant in samples with realistic thickness. 

Second, we put forward new insights into how the above-the-gap harmonic radiation is formed.
More specifically, we show that different parts of the HHG spectrum effectively originate 
from different depths below the sample surface. While the highest harmonics are generated
at or very close to the surface, the signal for the medium-order harmonics is sourced much deeper,
up to hundreds of nanometers. It is also shown here that the propagation and absorption effects
add significantly to the damping of higher harmonic orders.

\section{Evolution equations for pulsed electro-magnetic fields}

The description of the solid-state high-harmonic generation can be often treated 
in  the first Born approximation in the sense that one first calculates the evolution of the optical field
for the excitation pulse, including all necessary frequencies below the band-gap. The results are then
used to drive the high-harmonic generation for the frequency range above the material band gap.
While this is an approximation, it is an excellent one due to the fact that the intensity of the high harmonic radiation
is so low, and it is absorbed so quickly, that the feed-back effect on the propagation
of the driving pulse is truly negligible. Note that such a splitting of the simulation into a
``driver-'' and ``harmonic-stage'' has been utilized in the high-harmonic generation
modeling in gaseous media~\cite{Gaarde2008}, too.

\subsection{Excitation-pulse dynamics inside a slab-sample }

Consider a material sample characterized by its frequency-dependent permittivity $\epsilon(\omega)$,
in a plan-parallel slab geometry. The material is exposed to the excitation by 
an optical pulse incident on the ``entrance facet,'' not necessarily in the normal direction.
The Fresnel reflections from the material interfaces give rise to multiple pulses 
propagating inside the slab in ``forward'' and ``backward'' directions. For a very short
incident pulse and a sufficiently thick sample these counter-propagating waveforms do not
overlap or interact in the bulk of the material. However, the incident and reflected pulses always
overlap in the vicinity of the exit surface, where they create an interference pattern.
The interference between the  pulses
can be especially pronounced in materials with higher refractive indices, and the effect this has on
the harmonic generation is amplified by the extreme nonlinearity of the conversion process.
This  situation must be accurately captured by the simulation of the excitation pulse.

\smallskip

\noindent{\bf Propagation inside the bulk material:}\\
Away from the material boundaries, the displacement field can be decomposed into forward-
and backward-propagating components, 
\begin{equation}
\vec D = \vec D_+ + \vec D_- \ ,
\end{equation}
each expressed as a superposition of plane waves inside the slab,
\begin{equation}
\vec D_\pm(t,\vec r) = \int \vec A_\pm(t,\vec k) e^{\mp i  \omega(k) t} e^{i \vec k .\vec r}  d^3k \ .
\label{eqn:ansatz}
\end{equation}
Here $\omega(k)$ is treated as a function of the spatial frequency, and it satisfies the dispersion 
relation 
\begin{equation}
  k^2 = \omega(k)^2 \epsilon(\omega(k))/c^2 
  \ \ , \ \ k = \vert \vec k\vert \ .
\end{equation}
Since this relation defines $\omega(k)$ as an implicit function, the treatment must be restricted
to a single transparency window so that a unique complex-valued frequency can be assigned to a
given wavenumber $k$. Here we apply this framework to the below-gap frequencies to describe the
pump pulses, including their spectral broadening and below-gap harmonic generation.

In a complete analogy to the $t$-propagated unidirectional pulse propagation equation~\cite{tUPPE},
the vectorial spectral amplitudes that appear in (\ref{eqn:ansatz}) obey their individual propagation equations,
\begin{align}
  \partial_t \vec A_\pm(t,\vec k) &= \\
  \mp\frac{i \omega(k)}{2}&e^{\pm i \omega(k) t} \left[  \frac{k^2-\vec k \vec k}{k^2} . \vec P(\vec k)
    - \mu_0 \partial_t \vec J(\vec k) \right] . \nonumber
\end{align}
However, these evolution equations must be solved simultaneously because they are
coupled via the light-matter interaction in the slab, here expressed as
the induced  polarization $\vec P$ and current density $\vec J$
\begin{equation}
  \vec P(r,t) = \vec P(\{\vec E(r,t)\}) \ \ \text{and} \ \  \vec J(r,t) = \vec J(\{\vec E(r,t) \}) \  .
  \label{eqn:resp}
\end{equation}
%
Of course, the polarization calculation requires to first obtain the {\em total} electric field $E$ from
the {\em total} displacement field $D$, and this can be done by iterative inversion of
the constitutive material relation,
\begin{equation}
\vec E = \frac{1}{\epsilon}\left[ \vec D - \vec P(\{\vec E\}) \right] \ ,
\end{equation}  
as it is usual in the time-domain Maxwell-equations solvers. In a typical high-intensity femtosecond
pulse propagation in condensed media two or three iterations are sufficient. Given a model to calculate the
polarization and the current density for a given history of the electric field, the above equations
constitute an approximation-free evolution system for the Maxwell problem inside the sample.

\smallskip

\noindent{\bf Treatment of material interfaces:}\\
Besides the mutual coupling between the two $D_\pm$ inside the material wherever the forward- and
backward-propagating waves overlap, material interfaces
must be accounted for. Roughly speaking, the idea is to extend the spatial domain in (\ref{eqn:ansatz})
to the outside of the slab so that spectral propagator can be applied across the whole
sample, and also in the regions ``behind'' the material interfaces. Each of the two  $D_\pm$ amplitudes is
held by its dedicated ``computational domain'' in which the {\em linear} propagation is strictly
unidirectional (and governed by $\epsilon(\omega)$). These domains extend beyond the boundary of the
material slab by some fifty to hundred microns  
(along the beam direction, say $z$) in order to accommodate their ``input-'' and ``output-ports.''

The artificial output-region extending the computational domain outside the exit facet allows the spectral propagator
to evolve the optical wave-packet encountering the material interface beyond the material boundary.
However, the optical field amplitude outside of the sample is never used except to calculate the
spatial spectrum of the pulse via Fourier transform. The extended domain is eventually terminated
by an appodized absorbing region serving as a pulse dump.

On the ``entrance-side'' of the computational domain, the field behind the entrance facet must be constructed
anew before each application of the linear propagator as if the waveform propagated from a half-space space
filled by the same medium. However, the pre-calculated field in the ``input port'' must correctly account
for the reflection from the material boundary. This is achieved by a mapping of the field propagating
in the slab in the opposite direction, while applying the wave-number dependent Fresnel reflection coefficient,
$r(k_z,k_\perp)$, in the spectral representation of the pulse.

To illustrate the algorithm in a pseudo-code,
consider how the right-going field $D^+(z,r_\perp)$ partially reflects from the exit facet located at $z=z_0$
and is used to prepare the input port for the left-going field  $D^-(z,r_\perp)$ for locations
$z>z_0$, i.e. behind the material interface.
A single update is executed as follows:

\noindent
0. right-going $D^+(z,r_\perp)$ obtained in the previous step\\
1. calculate spatial spectrum $D^+(z,r_\perp)\rightarrow \hat D^+(k_z,k_\perp)$\\
2. apply reflection coefficient $r(k_z,k_\perp) \hat D^+(k_z,k_\perp)$\\
3. back to real space,  $D_r^+(z,r_\perp)\leftarrow r(k_z,k_\perp) \hat D^+(k_z,k_\perp)$\\
4. re-calculate  $D^-(z_0+z,r_\perp)=D_r^+(z_0-z,r_\perp)$ for $z>z_0$\\

Here, $\rightarrow$ and $\leftarrow$ stand for the transforms between the real-space and
spectral-space representations, each consisting of a Fourier transform connecting $z,k_z$ and
a Hankel transform connecting $r_\perp,k_\perp$.

This treatment of the internal reflections provides a numerically exact solution to the linear part
of the propagation problem, with the accuracy controlled by the extension of the spatial domain
along the pulse propagation direction. Of course, the spatial extension together with two additional
spectral transform make this algorithm expensive in comparison with unidirectional propagation.
However, this is a non-issue given that the calculations required to model the
material microscopic response (see Sect.IIIB) are orders of magnitude more expensive.

\subsection{Propagation of harmonic radiation}

Having calculated $\vec E(\vec r,t)$ inside the sample, we can use it in  point-by-point to
microscopic calculations to evaluate the induced current density $J(\vec r,t)$ which is responsible
for the above-the-gap harmonic radiation. This is done using the methods described in the subsequent
sections (Sect.IIIB). Here we give a description of how the source  $J(\vec r,t)$ converts into HHG radiation detected
outside of the sample.


In general, the HHG radiation detected outside of the sample is actually ``sourced'' in a sub-surface layer
several microns thick. Thus, the first step is to propagate the HHG radiation generated in depth of the
sample to its surface. In the second step, each spectral component is transmitted through the
material interface with a frequency-dependent Fresnel coefficient. Because of the highly nonlinear dependence
of the induced current density on the local amplitude of the driving field, the transverse profile
of the beam plays a role and it is crucial to account for it.

The propagation of the high-harmonic radiation through the sample is governed by the unidirectional
pulse propagation equation~\cite{gUPPE},
\begin{equation}
  \partial_z \vec S(z,\omega, k_\perp) =   i k_z S(z,\omega, k_\perp) 
  -\frac{\omega }{2 \epsilon_o c^2 k_z} J(z, \omega, k_\perp)  ,
\label{eqn:hhgprop}
\end{equation}
where the propagation constant
$k_z\equiv\sqrt{\omega^2/c^2 \epsilon(\omega) - k_\perp^2}$
reflects the absorption and dispersion properties for the above-the-gap frequencies
via $\epsilon(\omega)$, and $ J(z, \omega, k_\perp)$ is the spatial spectrum of
the current-density  induced by the driver field. In the first Born approximation, the
above is nothing  but a set of inhomogeneous first-order differential equations that
are straightforward to solve.

Because the beam spot is usually at least several $\mu$m in size, it is admissible to 
use a paraxial approximation for the short-wavelength HHG radiation. This amounts to
replacing the propagation constant by $k_z(\omega)=\omega n(\omega)/c$ in (\ref{eqn:hhgprop}).
Further, the spectrum is measured in the far field, and therefore the detected amplitude
we are interested in is $\vec S(z,\omega, k_\perp=0)$. The solution for the
detected spectral amplitude of HHG then reads,
\begin{equation}
   \vec S(z,\omega, 0) \sim
  \frac{\omega t(\omega) }{2 \epsilon_o c^2 k_z(\omega)} \int^z  \!\!\! e^{i k_z(\omega) (z-z')} J(z', \omega, 0) dz  ,
\label{eqn:hhgfarfield}
\end{equation}
where $t(\omega)$ stands for the Fresnel transmission coefficient (at normal incidence),
and where the on-axis far field amplitude of the induced current density can be calculated
as the integral over the beam cross section,
\begin{equation}
  J(z', \omega, 0) \sim \int J(z', \omega, r) r dr \ .
  \label{eqn:beam}
\end{equation}
This formula represents the coherent average over the excitation beam, and has a significant effect on
the resulting spectrum.

The most pronounced effect embodied in the above formulas is the frequency-dependent absorption
manifested  in propagator $e^{i k_z(\omega) (z-z')}$. It will be shown in the following that it is
the interplay between the HHG absorption and the spatially modulated source strength in $J(z', \omega, r)$
that controls which regions in the sample contribute most to the detected HHG spectrum.

\section{Material models}

This work utilizes a model of zinc blende materials for our simulation-based
illustration. In particular, we concentrate on GaAs,
because it is the material for which the propagation-effects in HHG were
studied in some detail~\cite{Xia:18}. We emphasize that nothing changes in
modeling a different material.

\subsection{Linear optical properties}

\begin{figure}[h!]
    \centerline{\includegraphics[clip,width=0.9\linewidth]{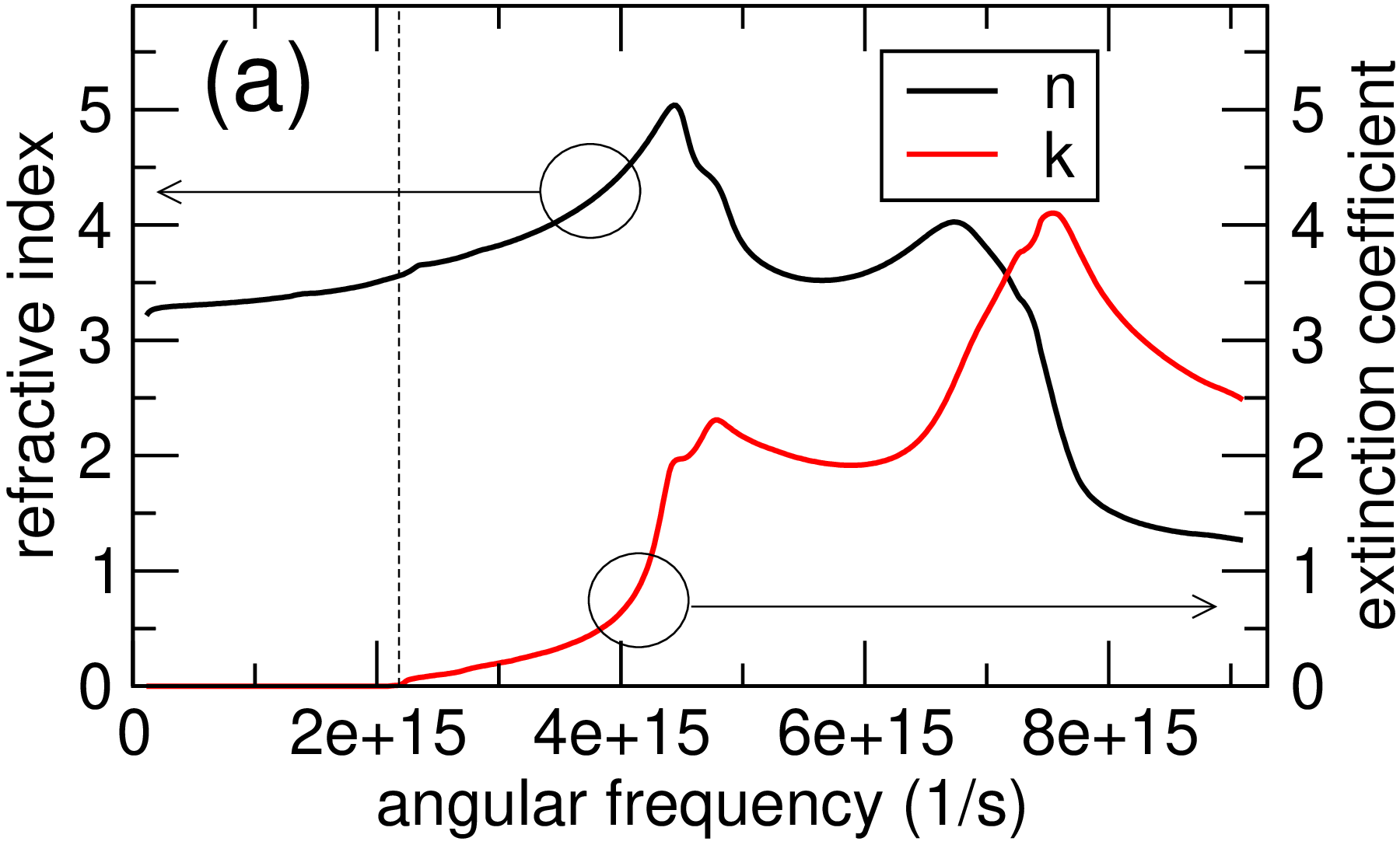}}
\caption{
  Refractive index and extinction coefficient in GaAs. 
\label{fig:GaAsNK}
  }
\end{figure}

The GaAs model for the linear chromatic properties was constructed by
joining three different sets of data, obtained from~\cite{refrinfo}
and originally published in~\cite{Aspnes,Skauli,Papatryfonos}.
Figure~\ref{fig:GaAsNK} illustrates the refractive index and extinction coefficient utilized for
the simulation of the propagation of the high-harmonic radiation.

This data is used to construct the frequency-dependent propagation constant $k_z(\omega)$ in the
paraxial-propagation formula~(\ref{eqn:hhgprop}). The tabulated $n(\omega)$ is also used
to calculate the frequency-dependent ``coupling pre-factor'' in front of the convolution integral.

\subsection{Nonlinear material response}

The nonlinear response of the material is calculated in the framework of Semiconductor Bloch Equations (SBE)~\cite{KochSBE},
which in turn requires a description of the electronic structure. For the band-structure we utilize the
semi-empirical tight-binding model~\cite{TBM1} with specific parameter sets representing
GaAs  adopted from~\cite{VOGL}. These material models were utilized and tested in the previous works,
demonstrating that the second-order nonlinear coefficient in GaAs~\cite{Kolesik:23} can be calculated,
and that measurements of HHG spectra were also reproduced without resorting to any parameter tuning~\cite{KolesikPRB}.
The sgiSBEs simulator, described in~\cite{AlgSBE}, calculates the time-dependent density matrix
$\rho_{mn}(\mathbf{k}; t)$ for each initial $\mathbf{k}$, and gives the induced current-density
as an integral over the Brillouin zone~\cite{Wilhelm21},
\be
\mathbf{j}(t) =
\sum_{mn}\int\displaylimits_\text{BZ} \text{Tr}\left[ \partial_{\kt} h(\kt) \rho(\kk ;t) \right] \frac{d\kk}{(2\pi)^3} \ ,
\label{eq:observeJ}
\ee
where the $\mathbf{k}$-dependent Hamiltonian $h(\mathbf{k})$ and its gradient is obtained in an explicit form from the
tight-binding model, and  $\mathbf{k}_t = \kk - \mathbf{A}(t)$ is slaved to the vector potential of the driving pulse.

The initial density matrix is set to represent all valence bands full
and the conduction bands empty. Each evolution step follows a scheme akin to the operator splitting, alternating
between the density matrix $\rho_\text{o}$ in the atomic-orbital basis, or $\rho_\text{h}$   in the instantaneous Hamiltonian basis
as follows:\\
0. $\rho_\text{o}(\mathbf{k},t-\Delta t)$ obtained in the previous step\\
1. diagonalization of $h(\kt)$ yields:\\
$\phantom{X}\bullet$ eigenvectors organized into matrix $V$\\
$\phantom{X}\bullet$ energies $\epsilon_a$ forming matrix $U = \text{diag}\{-i \Delta t\epsilon_a/\hbar\}$ \\
2. transformation to Hamiltonian basis,  $\rho_\text{h} \leftarrow V \rho_\text{o} V^\dagger$\\
3. evolution step, $\rho_\text{h} \leftarrow U \rho_\text{h} U^\dagger$\\
4. off-diagonal dephasing, $\rho_\text{h} \leftarrow \rho_\text{h}  \exp[-\Delta t/\tau_\text{dep}] $\\
5. transformation to orbital basis,  $\rho_\text{o} \leftarrow V^\dagger \rho_\text{h} V$\\
6. microscopic current, $j(\mathbf{k},t)\leftarrow\text{Tr}[\partial_{\kt} h(\kt) \rho_\text{o}(\kk;t)]$\\

While this implementation of the algorithm~\cite{AlgSBE} minimizes the number of matrix-matrix multiplication,
the overall numerical complexity is mainly given by the exact diagonalization step, and that is why ours is
a relatively expensive algorithm. The advantage of this approach is that the symmetry of the material is
faithfully preserved and that the noise-floor of the method is very low.

The phenomenological dephasing time $\tau_\text{dep}$ is chosen to be 5fs. For this relatively fast dephasing,
we can sample the Brillouin zone with a rather low resolution of $16^3$ points. While more realistic
results would require finer grid-resolutions, our conclusions would not change. In fact, it is safe to say
that the same conclusions can be reached with and apply to any HHG-model that can capture the symmetry of the
material.

\section{Results}

We assume that a Gaussian pulse, with a central wavelength of 3.5$\mu$m, has a peak intensity
of $10^9$V/m in the material just after entering the sample. The transverse size of the
collimated beam is chosen to be 0.3mm, so that the diffraction effects remain weak
during its propagation through material layers from tens to hundreds of microns.
The cross-section of the beam is sampled on a radial grid of 32 points (spaced suitably
for the discrete Hankel transform~\cite{DHT}).
The excitation pulse is polarized linearly along
the crystal axis, resulting in a geometry with a vanishing second-order nonlinearity.
The nonlinear light-matter interaction included in the driver simulation are the
instantaneous Kerr effect (with $n_2=1.2\times 10^{-17}$m$^2$/W, \cite{Hurlbut:07,Ensley:19}),
and the generation of excited carriers.
The latter is described by a phenomenological rate~\cite{Guide} scaling with $I^3$ and pre-factor
adjusted such that the ``ionization'' losses through a sample 650$\mu$m thick are about
fifty percent for the given initial peak amplitude, roughly matching the losses observed
in the experiment of Ref.~\cite{Xia:18}, which motivated our simulation setup.
Methods outlined in Sec.IIA are utilized for the first stage of the simulation
involving the excitation pulse.
We use spatial grids of up to $2^{18}$ points (along $z$) with a grid-spacing of 5nm 
and record the history of the electric field inside a 2$\mu$m
layer adjacent to the exit facet of the sample. The recording at each spatial point
used 8192 sampling points over a time-interval of 600~fs.

In the second stage of the simulation, we utilized the scheme described in Sec.IIB
to simulate the generated high-harmonic radiation and its propagation from the sample
towards its detection in the far field. The electric-field histories recorded in the
first stage were used as the excitation for the simulation of the material
response relevant for the above-the-gap frequency region as described in Sec.III.

\subsection{Propagation effects and space-time reshaping of the excitation pulse}

\begin{figure}[h!]
    \centerline{\includegraphics[clip,width=0.95\linewidth]{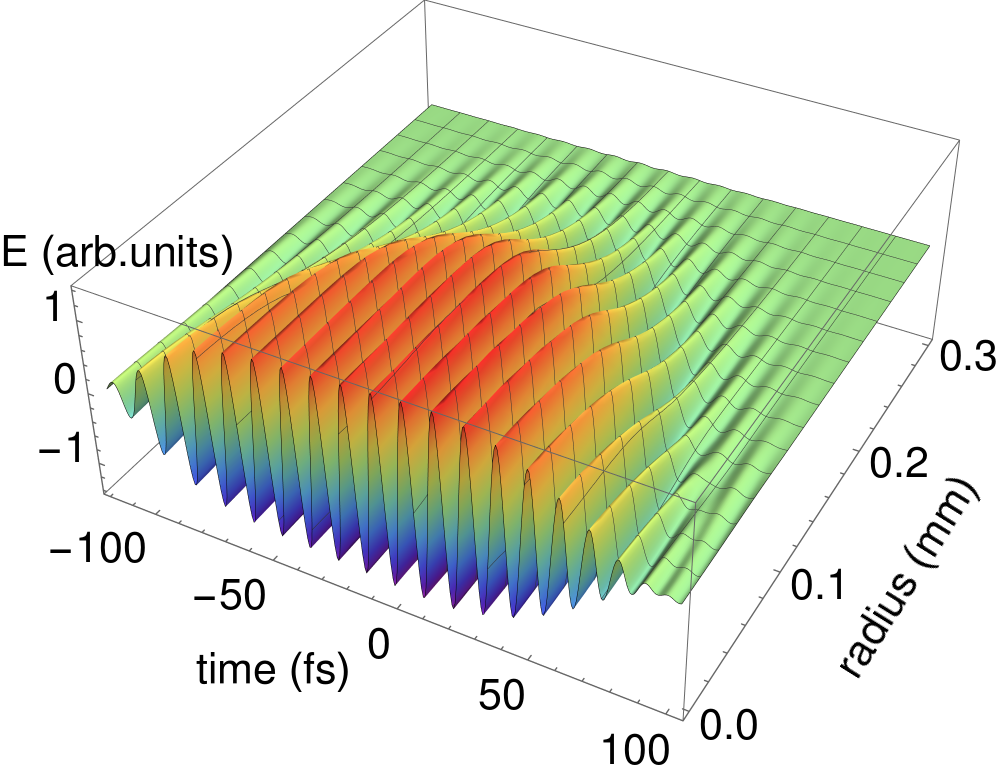}}
\caption{
  Spatio-temporal map of the electric field in the driving pulse, depicted after 650$\mu$m
  propagation through a GaAs sample.
  The color scheme was chosen to emphasize the flat-top shape in what
  was initially a Gaussian pulsed beam.
\label{fig:pulseTS}
  }
\end{figure}

Let us first illustrate the spatial-and-temporal reshaping the excitation pulse goes through.
Already after 50 microns of propagation, the pulse energy and amplitude of an originally
Gaussian pulse diminish significantly, 
and the waveform continues to reshape upon further propagation. Current carriers are mainly
generated in the high-intensity on-axis region, and cause a nonlinear phase shift which adds
curvature to the wave-fronts in the trailing portion of the pulse.
The amplitude reshaping is illustrated in Fig.~\ref{fig:pulseTS}, showing the driving pulse
resolved in radius and local time. Note that this ``snapshot'' is taken before the pulse reaches
the exit facet of the sample.
The color-scheme in the figure was chosen to
emphasize the most important feature, which is the flat-top peak amplitude.
This is a consequence of the losses induced by the carrier excitation, which affect
most the  highest-intensity parts of the waveform, and which effectively ``shaved-off'' the top
of the original Gaussian pulse.

\begin{figure}[h!]
    \centerline{\includegraphics[clip,width=0.95\linewidth]{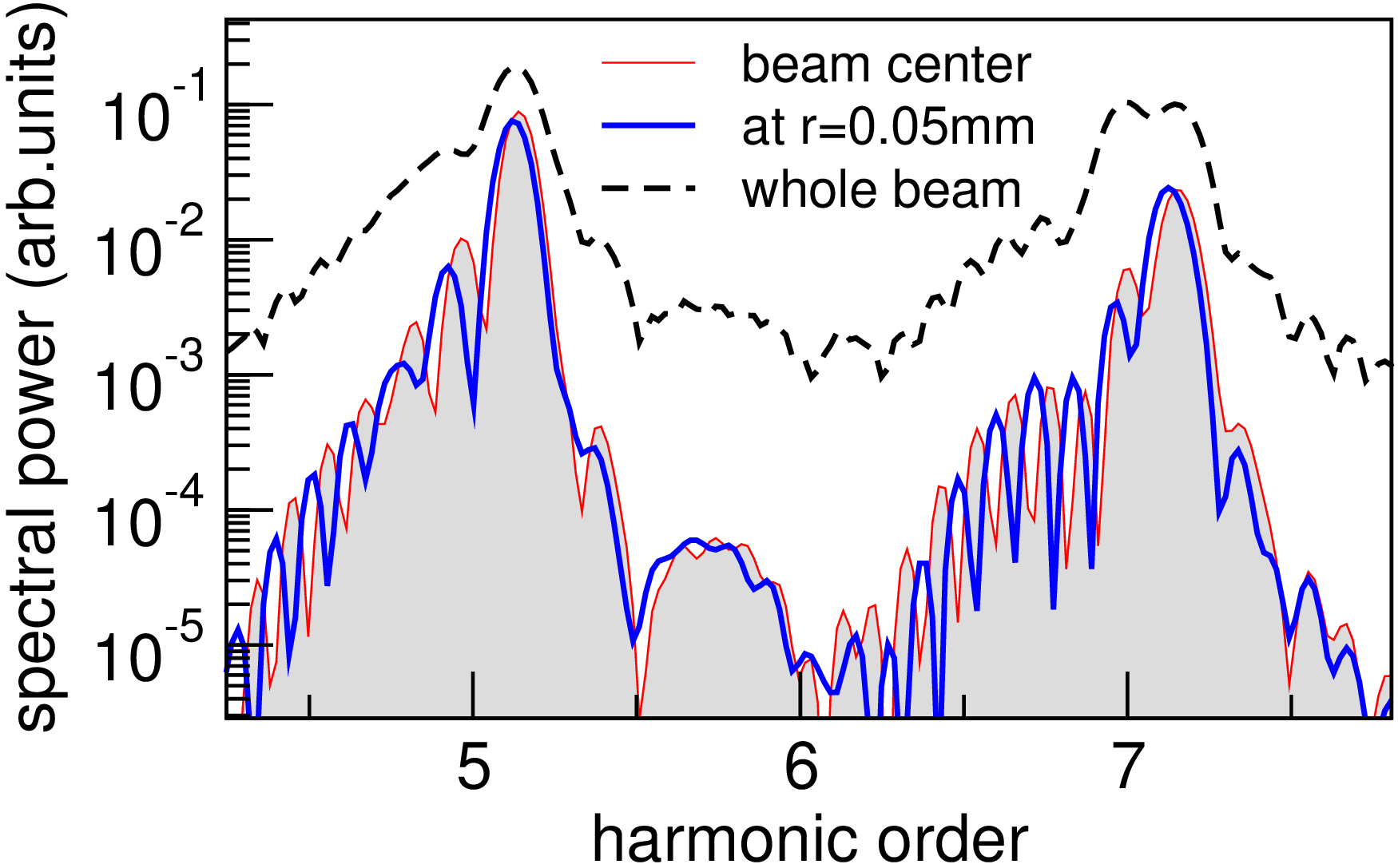}}
\caption{
  Spectral power of the induced current density exhibits modulations caused by the chirp
  of the driver pulse. The ``phase-shift'' between the modulations at different radial
  locations reflects different timing of the harmonic radiation at different radii.
\label{fig:spcvsrad}
  }
\end{figure}

The phase-shifts in the carrier wave modify the timing of the induced nonlinear current
density. In particular, the chirp gives a temporal shift between harmonics radiated by the
leading and trailing parts of the driver pulse. This shows up in the harmonic spectra
as a modulation, exemplified in Fig.~\ref{fig:spcvsrad}, where we show a detail of the spectral power
of the induced current density in the center of the beam compared to its counterpart at the
radial distance of 50 micron. While the two spectra exhibit the same power levels, their
modulation structures are shifted in relation to each other. This indicates that different
phase shift in the carrier at different radial distance cause different relative timing
between spectral components of the nonlinear current. These are the effects that a simple
intensity averaging~\cite{Multiscale} would not capture properly, and this is why we
calculate the microscopic response of the material from the actual radially-resolved
time-dependent electric fields.

\subsection{Interference effects at the exit facet}

Related to the propagation effects
is the Fresnel reflection of the driving pulse from the material interfaces.
Especially the internal reflection from the exit facet may be important because it can
significantly alter the peak amplitude of the driving pulse.

\begin{figure}[h!]
    \centerline{\includegraphics[clip,width=0.95\linewidth]{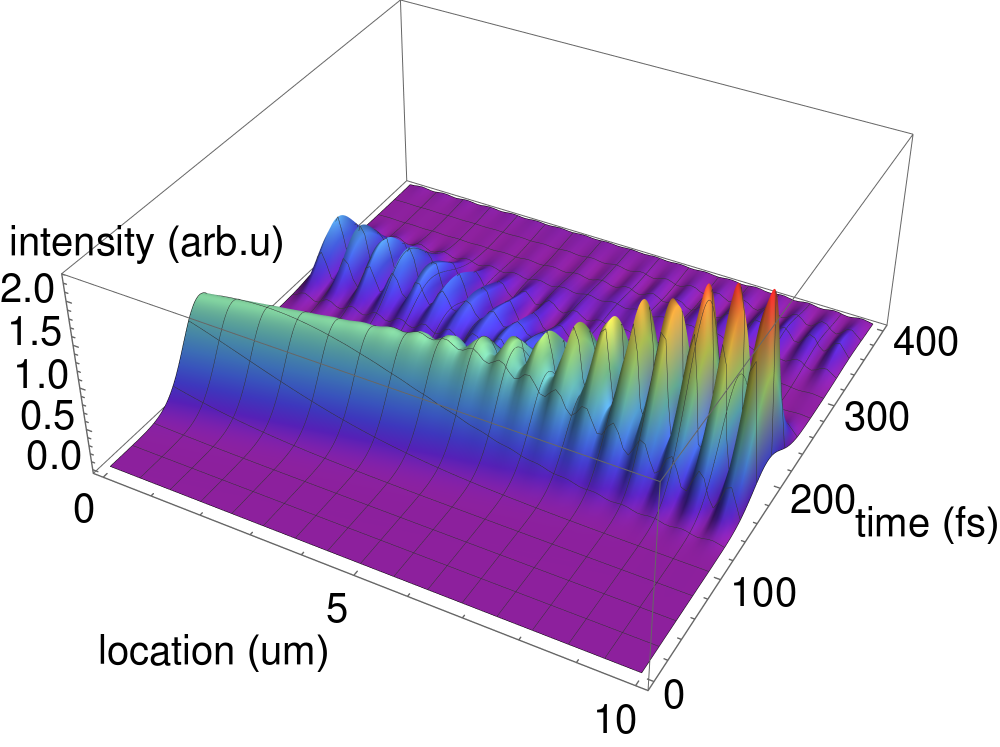}}
\caption{
  Cycle-averaged light intensity versus time and location in the sample of GaAs. Entrance and exit
  facets are located at the left-side edge of the figure (i.e. at $z=0\mu$m), and right at the location
  of right-most peak, respectively. Materials with high refractive indices, such as GaAs, give rise
  to a pronounced interference structure near the exit facet which in turn affects the nonlinear material
  response.
\label{fig:interference}
  }
\end{figure}

In the particular setting of this work, the reflections from the material interfaces play a major
role. The reason is illustrated in Fig.~\ref{fig:interference}, showing the inside of the material
slab 10$\mu$m thick, irradiated by a pulse entering from the left. Depicted here is the cycle-averaged
local intensity of the ``composite'' optical waveform which consists of pulses propagating forth and back
as in a Fabry-Perot cavity. 
%
The most important
feature is the interference pattern in the vicinity of the exit facet, where the local intensity reaches
considerably higher than in the incident pulse.
Here, the intensity profile is extended and drawn one quarter
wavelength beyond the exit facet of the sample in order to indicate the depth of the modulation, which is significant
due to the high refractive index of the material.

Given the extreme nonlinearity of the high-harmonic generation process,
the peak amplitude modulations in the vicinity of the exit sample-facet have significant effect
on the HHG spectrum, as we show next. While simulating the driver-pulse propagation in the GaAs
sample, we have recorded the time-dependent electric field within a layer adjacent to the exit facet.
Throughout this layer, the microscopic response of the material was calculated as described in Sec.IIIB,
giving the induced current density
$J(r,z,t)$, Fourier transform of which which enters as the source in the propagation equations~(\ref{eqn:hhgprop})
for the high-harmonic radiation.

\begin{figure}[t!]
    \centerline{\includegraphics[clip,width=0.8\linewidth]{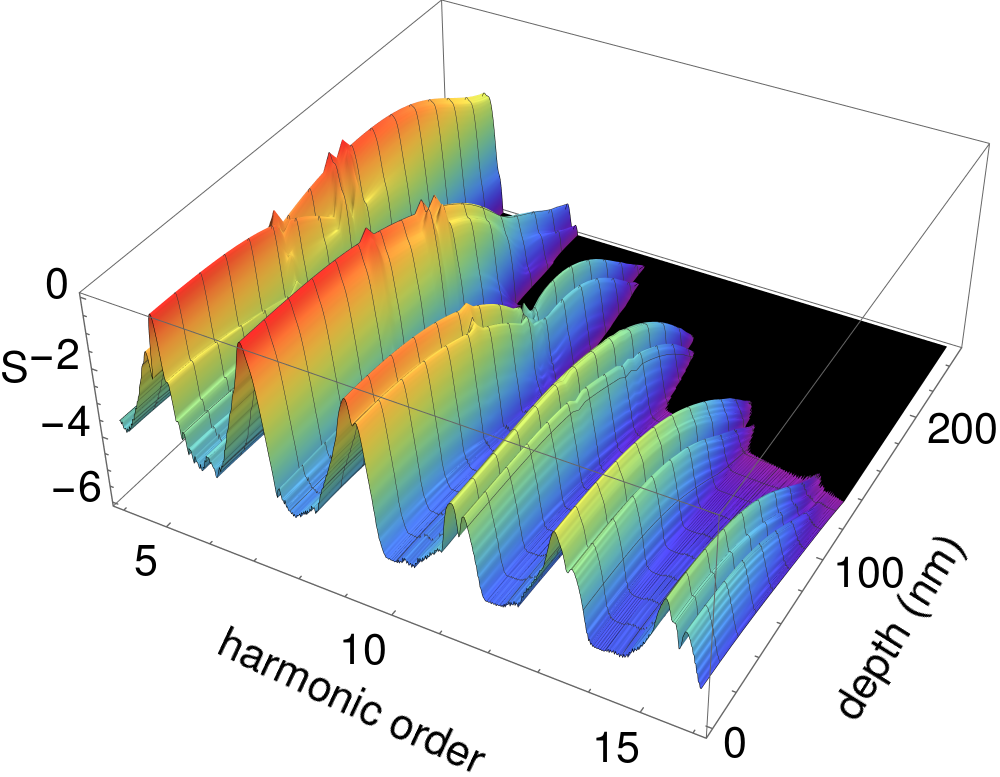}}
\caption{
  Logarithmic spectral power of the induced current density, $S=\log_{10}(|J(r=0,z,\omega)|^2)$ ,
  as a function of the depth  beneath the exit facet of the sample. 
  This example is for GaAs slab 170$\mu$m thick, excited  by a 60~fs pulse centered at $\lambda=3.5\mu$m.
\label{fig:Jdeep}
  }
\end{figure}

The logarithmic-scale spectral power of this source, $S=\log_{10}(|J(r=0,z,\omega)|^2)$, is shown in Fig.~\ref{fig:Jdeep}.
It is evident that as the local amplitude of the excitation field diminishes deeper from the sample surface,
the induced current exhibits orders of magnitude weaker high-frequency bands. This is a manifestation of the
interference fringe in the driving pulse localized at the sample facet. Of course, even deeper into the sample,
where the MIR intensity increases again in the second fringe, HHG spectra power also increases. However, due to
the absorption of the above-the-gap frequencies, those deeper regions give negligible contributions to the
observed HHG spectrum.

\begin{figure}[h!]
      \centerline{\includegraphics[clip,width=0.8\linewidth]{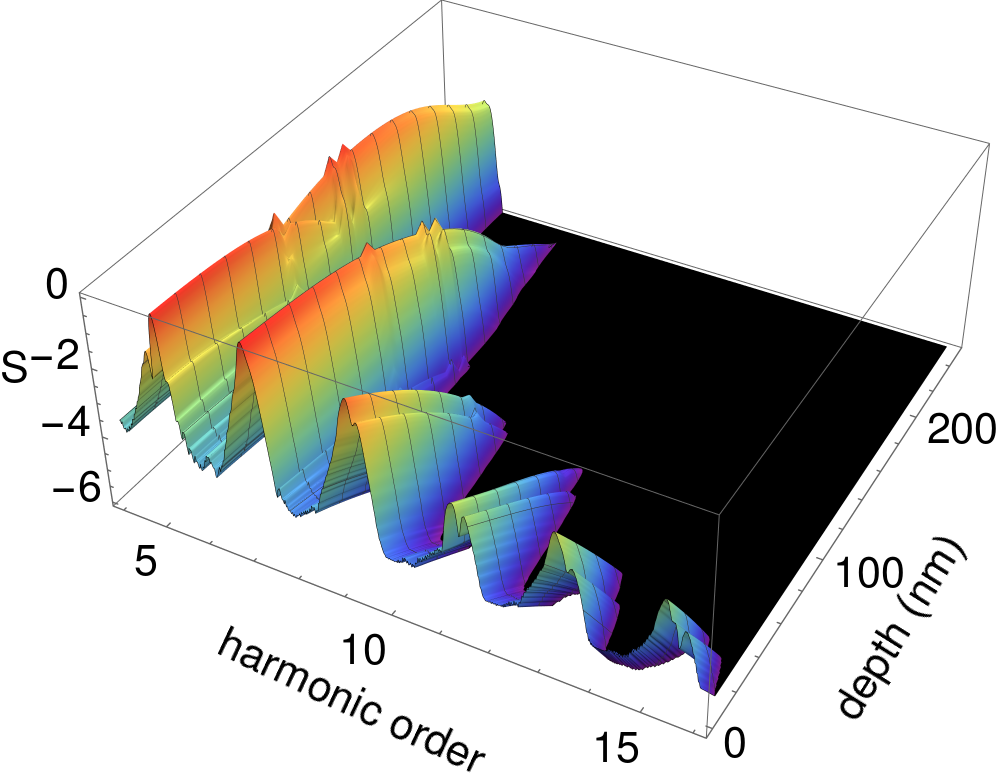}}
\caption{
  Spectral power of the high-harmonic generation detected at the exit facet, shown here as a function
  of the depth in the sample where the radiation was emitted. Contributions from the deeper regions
  are effectively extinguished in the highest harmonic bands.
\label{fig:Jsurf}
  }
\end{figure}

\subsection{Propagation and absorption of the high-harmonic radiation}

In HHG simulations, the spectral power of the induced current is often assumed to approximate the actual
observed HHG spectrum, essentially neglecting the difference between a source term in the Maxwell equations
and the radiation the term generates. In this section we show that it is an utterly unrealistic assumption
for the transmission geometry where the propagation-effects shape the spectral content of the high-harmonic
radiation. 

The relevant relation between the source term $J(\omega)$ and the spectrum of the detected radiation, $S(\omega)$,
is expressed in Eq.~(\ref{eqn:hhgfarfield}) which accounts for the propagation
and absorption of HHG radiation from its source to  outside of the sample. To appreciate the absorption
and its effect on the reshaping of the HHG-spectrum, Fig.~\ref{fig:Jsurf} depicts the spectral power sourced at
a given depth as it arrives at the sample surface.  More precisely, the quantity shown is
$S = \log_{10}( |e^{i k_z(\omega) z} J(z, \omega,0)|^2)$, so that the effect of the absorption can be appreciated.

The contrast with Fig.~\ref{fig:Jdeep} is indeed stark, especially for higher harmonic orders. Only a
very thin surface layer contributes to the highest observed orders, and this is obviously
due to the fact that a few hundred nano-meters of the material suffices to attenuate the radiation that
was generated deeper in the sample. 
In contrast, 
the lower above-the-gap harmonics carry contributions from a much thicker layer of the material.

\subsection{HHG spectra observed outside the material sample}

While the previous illustration shows the frequency-dependent  attenuation effect of the material
between the origin of the radiation and the surface, it says little about the actually detected spectrum.
To calculate this, formula (\ref{eqn:hhgfarfield}) calls for a ``coherent sum'' of all contributions from the
depths of the sample.

\begin{figure}[h!]
  \centerline{\includegraphics[clip,width=0.8\linewidth]{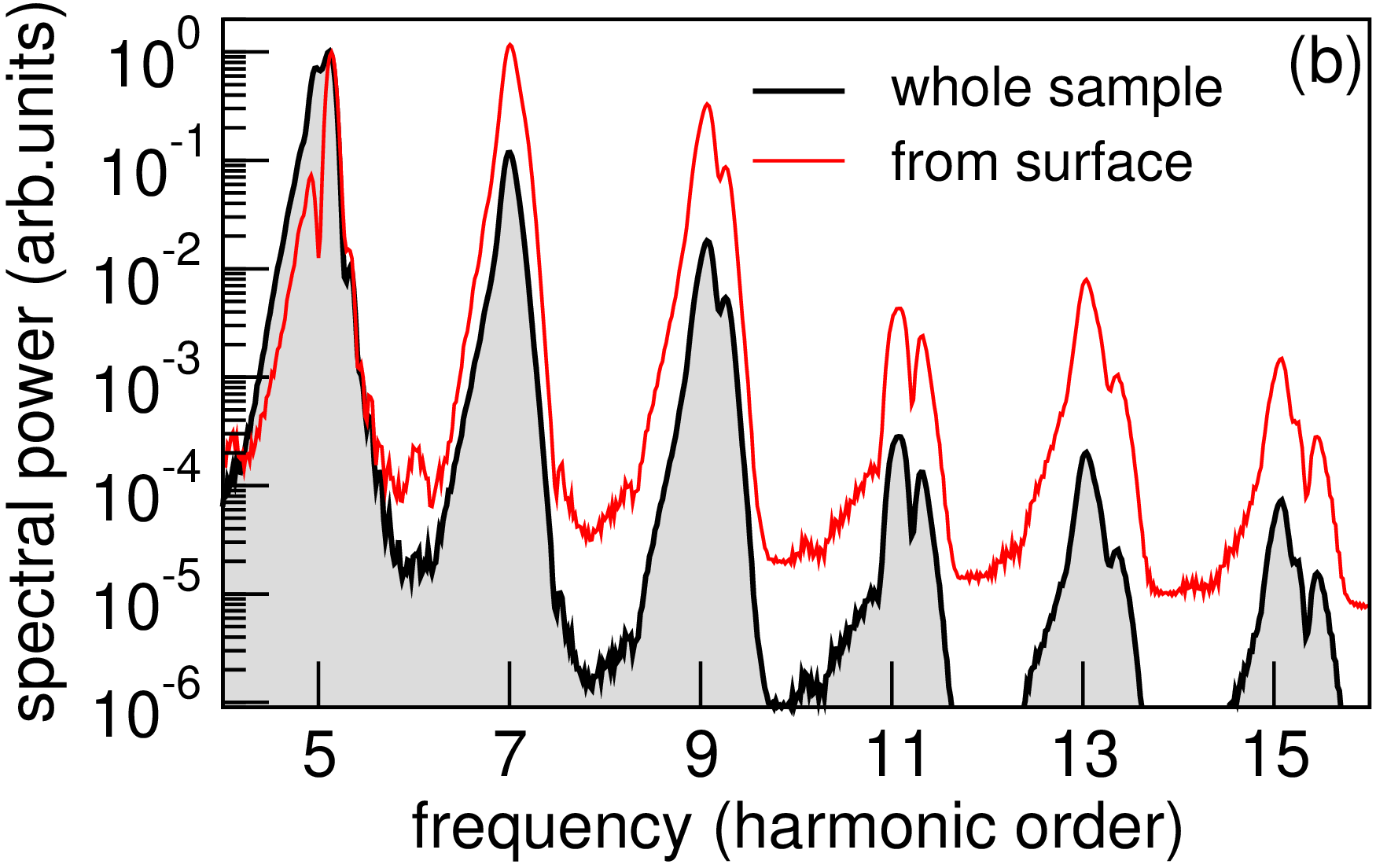}}
  \centerline{\includegraphics[clip,width=0.8\linewidth]{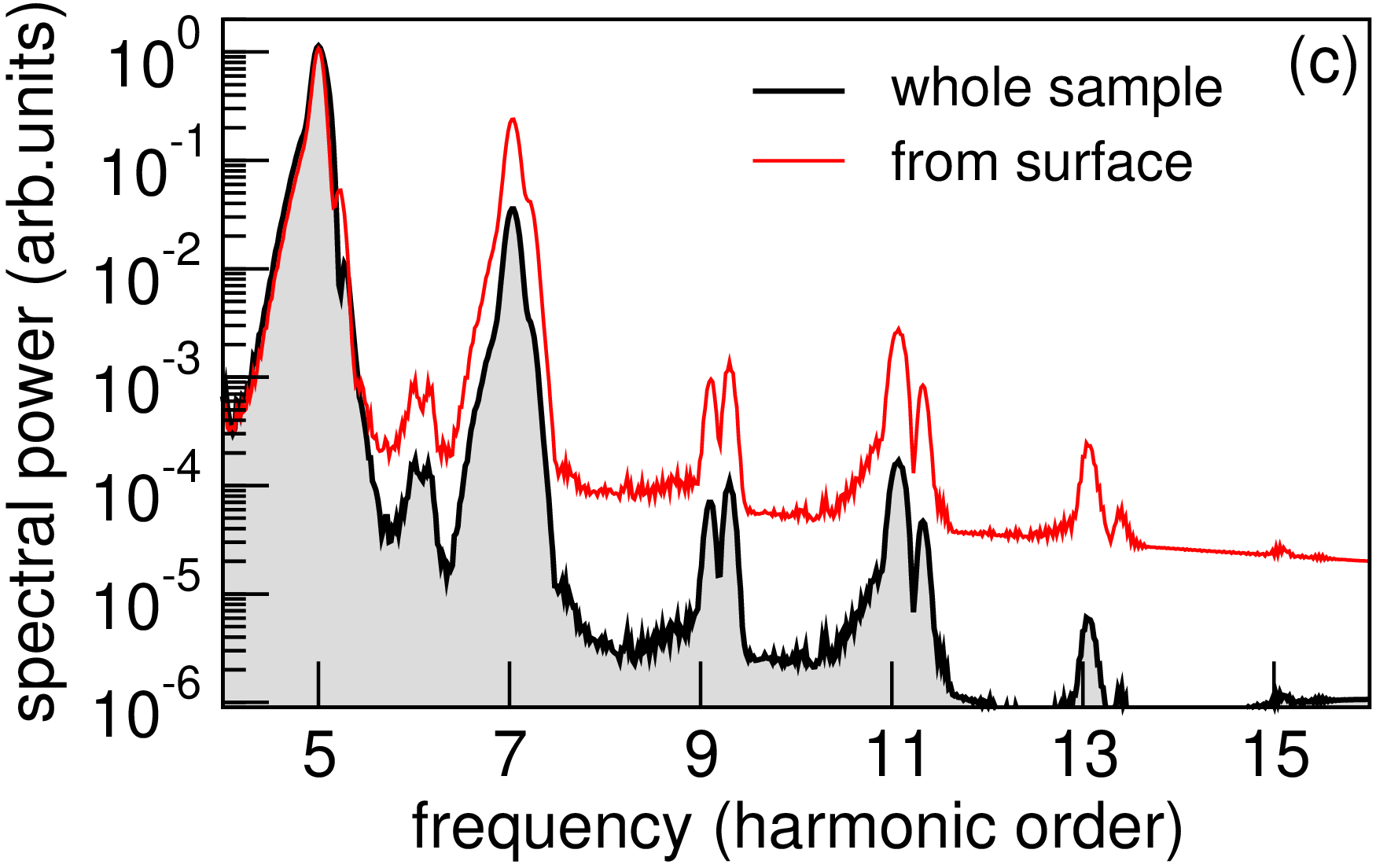}}
  \centerline{\includegraphics[clip,width=0.8\linewidth]{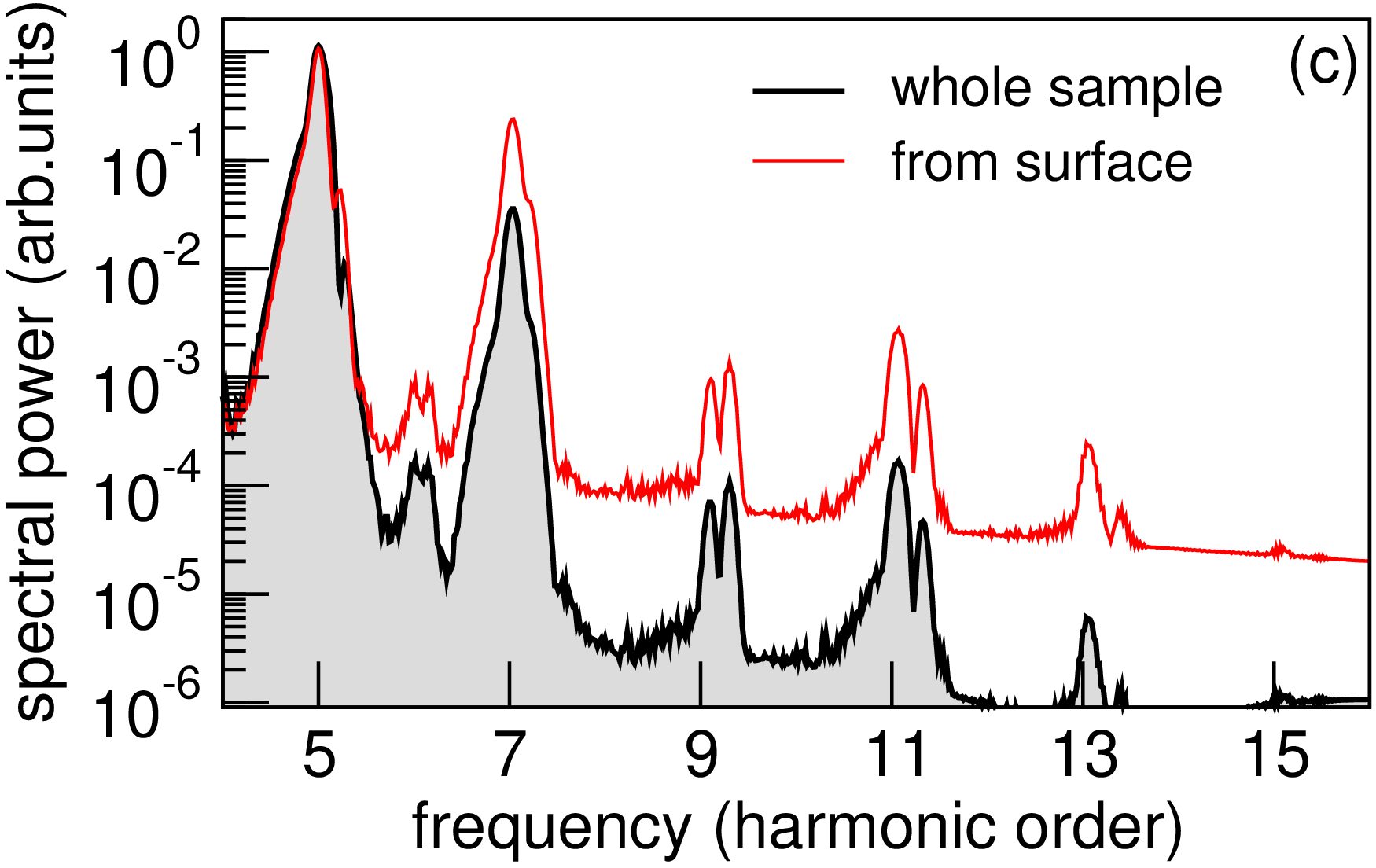}}
\caption{
  The effect of the propagation and absorption in the above-gap harmonic radiation
  for (a) 45, (b) 170, and (c) 650 $\mu$m thick material sample. HHG-spectra obtained solely
  from the surface overestimate the higher harmonic bands. This ``damping effect'' is in
  addition to that due to the nonlinear absorption of the pump.
\label{fig:obs}
  }
\end{figure}

Figure~\ref{fig:obs} shows the results for three different sample thicknesses, namely 45, 170, and 650 microns.
The spectra measured in the far field outside the sample are shown as the gray-shaded area below the thick
black line. In a qualitative agreement with the experimental observations in Ref.~\cite{Xia:18},
it is evident that the thicker samples effectively suppress the higher harmonic orders. While the main reason
for this was identified as due to the propagation effects reshaping and attenuating the MIR driver pulse,
here we show that that the propagation effects experienced by the high-harmonic radiation itself are equally
important. To make this point, each panel also
depicts a HHG spectrum (top, red line) calculated from the naive assumption that the observed signal is
dominated by its strongest component originating right at the sample surface. In other words, such a calculation
accounts for the propagation effects of the driver pulse, but neglects the propagation effects acting on
the high-harmonic radiation. As one can see, neglecting the latter gives spectra that overestimate the spectral
power of the higher-harmonic bands by more than an order of magnitude.

\section{Conclusions}

We have presented a framework  integrating realistic HHG-simulations into a comprehensive model of an experiment
in the transmission geometry, where the propagation effects experienced by the excitation pulse influence the
measurements in a crucial way. Our approach represents a numerically less demanding alternative to the 
time-domain DFT simulations of the material coupled with the one-dimensional wave equation for the electromagnetic
field. An important advantage of the method we put forward is the ability to simulate realistically thick samples
while including a full-resolution 3D description of the pulsed beam which is necessary to capture its spatial and
temporal reshaping.

Beyond the previously discussed propagation effects, namely the nonlinear absorption and self-phase modulation
of the excitation pulse, we have identified several additional mechanisms that are equally important or perhaps
even more consequential in shaping the HHG spectra measured in the transmission-geometry.

The first effect is the {\em spatial-temporal} reshaping of the excitation pulse which influences how different
parts of the cross-section in the pulsed beam contribute to the measured spectrum. As the excitation waveform
propagates through the material, it excites current carriers which in turn impart time- and radially-dependent
nonlinear phase-shifts. As a consequence, the high harmonics generated at different radii emerge with
different timings and varying chirps, giving rise to ``point-spectra'' with pronounced modulations structures.
The measured far-field spectrum, being a coherent sum over the cross-section of the beam, appears much smoother.

Second, we have shown that the interference effects created by the Fresnel reflections at the output facet
of the sample substantially change the peak amplitude, and this has a pronounced effect on the resulting HHG
spectrum. Due to the  extreme nonlinear nature of the HHG process, neglecting the role of the material interface
underestimates the relevant peak amplitudes and gives much weaker high-harmonic spectra. 

Third, and the most important is the propagation effect in the HHG radiation itself. As the HHG radiation propagates
through the material before it exits from the sample, the absorption adds significantly
to the damping of the higher-order harmonic bands.  These propagation effects together with the spatial
modulation of the nonlinear material response decide  which region of the sample contributes most to a given portion
of the spectrum. While the highest orders are generated only in a very thin surface layer, medium-order
harmonics are sourced from the deeper regions of of the material. An important implication here is that a
``point-model response'' simulated at a single spatial location does not provide a good approximation of the actual HHG spectrum.

The main take-away from our study is that no quantitative or even semi-quantitative
interpretation of the HHG spectrum taken in the transmission geometry is possible without
a comprehensive modeling which includes the (3D+1) simulation of the excitation pulse,
sample geometry including Fresnel reflections from its facets, and the propagation of the
HHG radiation through the lossy material. However, our work also demonstrates that thanks to the natural
separation between the dynamics of the below- and above-the-gap frequencies, it possible to construct
realistic HHG models at an acceptable computational cost. While many of the HHG experiments with
solid-state materials are done in the transmission geometry, accurate simulations should facilitate
their interpretation.

\section*{Acknowledgments}

Authors acknowledge the support from the
Air Force Office for Scientific Research under grants no.
FA9550-22-1-0182
and
FA9550-21-1-0463.


%

\end{document}